\documentclass[11pt]{article}
\usepackage{amsmath}
\usepackage{amsfonts}

\input{tcilatex}
\input{tcilatex}
\begin{document}

\begin{center}
\textbf{A NEW PERSPECTIVE ON NEWTON'S LAW OF COOLING IN FRAME OF NEWLY
DEFINED FRACTIONAL CONFORMABLE DERIVATIVE}

\bigskip

Erdal BAS$^{1,a}$, Ramazan OZARSLAN$^{1,b}$, Ahu ERCAN$^{1,c}$

\bigskip

$^{1}$\textit{Department of Mathematics, Faculty of Science, Firat
University, Elazig, 23119, Turkey}

\bigskip

email: $^{a}$erdalmat@yahoo.com, \textit{\ }$^{b}$\textit{%
ozarslanramazan@gmail.com, }$^{c}$ahuduman24@gmail.com

\bigskip
\end{center}

\noindent \textbf{Abstract. }In this paper, Newton's law of cooling is
considered from a different perspective with newly defined fractional
conformable. Obtained results are compared with experimental results and
found optimal fractional orders which fit better with real data. Results
show that Newton's law of cooling with fractional conformable derivative
gives better results to integer order derivative. Results are given
comparatively to Newton's law of cooling with integer order and experimental
data and also, fractional conformable derivative's advantages are supported
by numerical illustrations and error analysis.

\bigskip


\bigskip

\noindent \textbf{PACS 2010}: 47.54.Bd, 47.54.De,45.10.Hj

\section{\textbf{Introduction }}

Fractional differential equations idea was firstly suggested by Leibniz on
generalizing of integer order derivative three centuries ago. Many problems
with real world phenomena were modelled by integer order derivatives. If the
variable in the integer order derivative is specified as time, it gives the
amount of change according to time and so, approximate amount of change can
be calculated by changing time. Fractional derivative definitions involve
variable order and hence, it is not hard to estimate to give better
approximations to the integer order derivative. However, many fractional
derivative definitions have been introduced in recent years. These new
definitions have some advantages and disadvantages, for example, fractional
differential problems with Riemann-Liouville derivative do not include
initial conditions with integer order, but fractional differential problems
with Caputo derivative include initial conditions with integer order.
Therefore, the Caputo derivative is more useful than the Riemann-Liouville
derivative in some engineering and physical problems. Fractional derivatives
with exponential and Mittag-Leffler kernel have been introduced by
respectively Caputo-Fabrizio \cite{cf} and Atangana-Baleanu \cite{ab}. These
new operators are more useful in some real world problems due to having
non-singularity in its kernels.

Conformable derivative definition was firstly given by Khalil et al. \cite%
{khalil, ander}, and this operator shows similarity to the integer order
derivative, differently it includes a shift as $\varepsilon t^{-\alpha }$ in
its limit definition, but it is not conformable at $\alpha =0,$ i.e. $\lim
\limits_{\alpha \rightarrow 0}$ $_{a}D_{t}^{\alpha }f\neq f.$ More recently,
a new fractional version of conformable derivative definition has been
introduced by Jarad et al. \cite{jarad} in Riemann-Liouville and Caputo
sense. Lately, Atangana et al. have introduced $\beta -$derivative and
motivated by this, Morales et al. \cite{gomez} have introduced fractional
conformable $\beta -$derivative.

Newton's law of cooling gives the approximate amount of change in
temperature of cooling or heating body to ambient temperature. Elizabeth
Gieseking \cite{eliza} tested experimentally Newton's law of cooling for
three beakers of water for $100$, $300$ and $800$ $ml$ volumes. Gieseking
measured temperature of beakers every minute and compared real data to
classical Newton's law of cooling. More recently, Mondol et al. \cite{mondol}
have considered Newton's law of cooling with Riemann-Liouville and Caputo
sense and have tested experimentally for different liquids. They have
studied to obtain to fit real data by using fractional derivatives in
Riemann-Liouville and Caputo sense. Ortega et al. \cite{ortega} have
considered Newton's law of cooling with conformable derivative. Almeida et
al. \cite{almeida, almeida2} have used some new fractional derivatives to
obtain better results to fit data for some modelling problems. Also, some
physical modelling problems are studied by \cite{haci, haci2}.

In this paper, we consider Newton's law of cooling from a different
perspective with newly defined fractional conformable and $\beta -$%
derivatives. Firstly, we obtain new analytical solutions for Newton's law of
cooling with newly defined fractional conformable and $\beta -$derivatives,
and then we compare these new analytical solutions with experimental
results, tested by Gieseking \cite{eliza}, and study to find optimal
fractional orders which fit better with real data. Results show that
Newton's law of cooling with fractional conformable derivative gives better
results to integer order derivative. Results are given comparatively to
Newton's law of cooling with integer order and experimental data and also,
fractional conformable derivative's advantages are supported by numerical
simulations and error analysis.

\section{\textbf{Preliminaries}}

\noindent \textbf{Definition 1.} \cite{podlub} The Riemann-Liouville
derivative of order $\alpha $ is defined as 
\begin{equation*}
_{a}^{RL}D_{t}^{\alpha }f\left( t\right) =\frac{1}{\Gamma \left( n-\alpha
\right) }\frac{d^{n}}{dt^{n}}\dint \limits_{a}^{t}f\left( \xi \right) \left(
t-\xi \right) ^{n-\alpha -1}d\xi ,\text{ \ }n-1<\alpha <n.
\end{equation*}

\bigskip

\noindent \textbf{Definition 2. \cite{podlub} }The Liouville-Caputo
derivative of order $\alpha $ is defined as%
\begin{equation*}
_{a}^{C}D_{t}^{\alpha }f\left( t\right) =\frac{1}{\Gamma \left( n-\alpha
\right) }\dint \limits_{a}^{t}\frac{d^{n}}{d\xi ^{n}}f\left( \xi \right)
\left( t-\xi \right) ^{n-\alpha -1}d\xi ,\text{ \ }n-1<\alpha <n.
\end{equation*}

\bigskip

\noindent \textbf{Definition 3.} \cite{cf} The fractional derivative with
exponential kernel for $\alpha >0$ is defined as%
\begin{equation*}
_{a}^{CFC}D_{t}^{\alpha }f\left( t\right) =\frac{M\left( \alpha \right) }{%
n-\alpha }\dint \limits_{a}^{t}\frac{d^{n}}{d\xi ^{n}}f\left( \xi \right)
e^{-\frac{\alpha }{n-\alpha }\left( t-\xi \right) }d\xi ,\text{ \ }%
n-1<\alpha <n,
\end{equation*}%
where $M\left( \alpha \right) $ is a normalization constant that depends of $%
\alpha ,$ which satisfies that, $M\left( 0\right) =M\left( 1\right) =1.$

\bigskip

\noindent \textbf{Definition 4.} \cite{ab} The fractional derivative with
Mittag-Leffler kernel is defined as%
\begin{equation*}
_{a}^{ABC}D_{t}^{\alpha }f\left( t\right) =\frac{B\left( \alpha \right) }{%
n-\alpha }\dint \limits_{a}^{t}\frac{d^{n}}{d\xi ^{n}}f\left( \xi \right)
E_{\alpha }\left( -\frac{\alpha }{n-\alpha }\left( t-\xi \right) ^{\alpha
}\right) d\xi ,\text{ \ }n-1<\alpha <n
\end{equation*}%
where $B\left( \alpha \right) $ is normalization function and $B\left(
0\right) =B\left( 1\right) =1.$

\bigskip

\noindent \textbf{Definition 5. \cite{ab}.} Let $f:\left[ a,\infty \right)
\rightarrow 
\mathbb{R}
.$ The conformable derivative of $f\left( t\right) $ is defined as follows%
\begin{equation*}
_{a}D_{t}^{\alpha }f\left( t\right) =\lim_{\varepsilon \rightarrow 0}\frac{%
f\left( t+\varepsilon t^{1-\alpha }\right) -f\left( t\right) }{\varepsilon }
\end{equation*}%
for all $t>0,$ $\alpha \in \left( 0,1\right] .$ If $f\left( t\right) $ is $%
\alpha -$differentiable in some $\left( 0,a\right) ,$ $a>0$ and $\lim
\limits_{\varepsilon \rightarrow 0^{+}}f^{\left( \alpha \right) }\left(
t\right) $ exist, then define $\lim \limits_{t\rightarrow 0^{+}}f^{\left(
\alpha \right) }\left( t\right) =f^{\left( \alpha \right) }\left( 0\right) .$

\bigskip

\noindent \textbf{Definition 6.} \cite{thabet, khalil} The left and right
conformable integrals are defined as 
\begin{equation}
_{a}I^{\alpha }f\left( x\right) =\dint \limits_{a}^{x}\left( t-a\right)
^{\alpha -1}f\left( t\right) dt,\text{ \ }x\geq a,\text{ }0<\alpha \leq 1 
\tag{1}  \label{a5}
\end{equation}%
\begin{equation*}
I_{b}^{\alpha }f\left( x\right) =\dint \limits_{x}^{b}\left( b-t\right)
^{\alpha -1}f\left( t\right) dt,\text{ \ }x\leq b.
\end{equation*}

\bigskip

\noindent \textbf{Definition 7. }\cite{jarad} Fractional conformable
integral is defined as\textbf{\ }%
\begin{equation}
_{a}^{\beta }I^{\alpha }f\left( x\right) =\frac{1}{\Gamma \left( \beta
\right) }\dint \limits_{a}^{x}\left( \frac{\left( x-a\right) ^{\alpha
}-\left( t-a\right) ^{\alpha }}{\alpha }\right) ^{\beta -1}\frac{f\left(
t\right) }{\left( t-a\right) ^{1-\alpha }}dt  \tag{2}  \label{a6}
\end{equation}

\bigskip

\noindent \textbf{Theorem 1.} \cite{jarad} Let $\func{Re}\left( \beta
\right) \geq 0,$ $n=\left[ \func{Re}\left( \beta \right) \right] +1,$ $f\in
C_{\alpha ,a}^{n}\left( \left[ a,b\right] \right) .$ Then, Riemann-Liouville
fractional conformable derivatives are defined as follows, 
\begin{equation}
_{a}^{\beta }D^{\alpha }f\left( x\right) =\frac{_{a}D^{\alpha }f\left(
t\right) }{\Gamma \left( n-\beta \right) }\dint \limits_{a}^{x}\left( \frac{%
\left( x-a\right) ^{\alpha }-\left( t-a\right) ^{\alpha }}{\alpha }\right)
^{n-\beta -1}\frac{f\left( t\right) }{\left( t-a\right) ^{1-\alpha }}dt, 
\tag{3}  \label{a7}
\end{equation}%
and 
\begin{equation}
^{\beta }D_{b}^{\alpha }f\left( x\right) =\frac{\left( -1\right)
^{n}D_{b}^{\alpha }f\left( t\right) }{\Gamma \left( n-\beta \right) }\dint
\limits_{x}^{b}\left( \frac{\left( b-x\right) ^{\alpha }-\left( b-t\right)
^{\alpha }}{\alpha }\right) ^{n-\beta -1}\frac{f\left( t\right) }{\left(
b-t\right) ^{1-\alpha }}dt.  \tag{4}  \label{a8}
\end{equation}

\bigskip

\noindent \textbf{Theorem 2.} \cite{jarad} Let $\func{Re}\left( \beta
\right) \geq 0,$ $n=\left[ \func{Re}\left( \beta \right) \right] +1,$ $f\in
C_{\alpha ,a}^{n}\left( \left[ a,b\right] \right) .$ Then Caputo fractional
conformable derivatives are given by 
\begin{equation}
_{a}^{C\beta }D^{\alpha }f\left( x\right) =\frac{1}{\Gamma \left( n-\beta
\right) }\dint \limits_{a}^{x}\left( \frac{\left( x-a\right) ^{\alpha
}-\left( t-a\right) ^{\alpha }}{\alpha }\right) ^{n-\beta -1}\frac{%
_{a}D^{\alpha }f\left( t\right) }{\left( t-a\right) ^{1-\alpha }}dt,  \tag{5}
\label{a9}
\end{equation}%
and 
\begin{equation}
^{C\beta }D_{b}^{\alpha }f\left( x\right) =\frac{\left( -1\right) ^{n}}{%
\Gamma \left( n-\beta \right) }\dint \limits_{x}^{b}\left( \frac{\left(
b-x\right) ^{\alpha }-\left( b-t\right) ^{\alpha }}{\alpha }\right)
^{n-\beta -1}\frac{D_{b}^{\alpha }f\left( t\right) }{\left( b-t\right)
^{1-\alpha }}dt.  \tag{6}  \label{a10}
\end{equation}

\bigskip

\noindent \textbf{Definition 8. }\cite{abdon} Let $f:\left[ -\frac{a}{\Gamma
\left( \alpha \right) },\infty \right) \rightarrow 
\mathbb{R}
,$ then a different type of conformable derivative of $f\left( t\right) $ is
defined as%
\begin{equation*}
_{0}^{A}D^{\alpha }f\left( x\right) =\lim_{\varepsilon \rightarrow 0}\frac{%
f\left( t+\varepsilon \left( t+\frac{1}{\Gamma \left( \alpha \right) }%
\right) ^{1-\alpha }\right) -f\left( t\right) }{\varepsilon }.
\end{equation*}%
The different type of left conformable integral is defined as%
\begin{equation*}
_{0}^{A}I^{\alpha }f\left( x\right) =\int \limits_{0}^{t}\frac{f\left(
x\right) }{\left( x+\frac{1}{\Gamma \left( \alpha \right) }\right)
^{1-\alpha }},0<\alpha \leq 1.
\end{equation*}

\bigskip

\noindent \textbf{Theorem 3.} \cite{gomez} Let $\func{Re}\left( \beta
\right) \geq 0,$ $n=\left[ \func{Re}\left( \beta \right) \right] +1,$ $f\in
C_{\alpha ,a}^{n}\left( \left[ a,b\right] \right) .$ Then a different type
of Caputo fractional conformable derivatives are defined as follows, 
\begin{equation}
_{a}^{AC\beta }D^{\alpha }f\left( x\right) =\frac{1}{\Gamma \left( n-\beta
\right) }\dint \limits_{a}^{x}\left( \frac{\left( x+\frac{a}{\Gamma \left(
\alpha \right) }\right) ^{\alpha }-\left( t+\frac{a}{\Gamma \left( \alpha
\right) }\right) ^{\alpha }}{\alpha }\right) ^{n-\beta -1}\frac{_{\quad
a}^{A\quad n}D_{t-}^{\alpha }f\left( t\right) }{\left( t+\frac{a}{\Gamma
\left( \alpha \right) }\right) ^{1-\alpha }}dt,  \tag{7}  \label{a11}
\end{equation}%
and 
\begin{equation}
^{AC\beta }D_{b}^{\alpha }f\left( x\right) =\frac{\left( -1\right) ^{n}}{%
\Gamma \left( n-\beta \right) }\dint \limits_{x}^{b}\left( \frac{\left( 
\frac{b}{\Gamma \left( \alpha \right) }+t\right) ^{\alpha }-\left( \frac{b}{%
\Gamma \left( \alpha \right) }+x\right) ^{\alpha }}{\alpha }\right)
^{n-\beta -1}\frac{_{\quad t}^{A\quad n}D_{b-}^{\alpha }f\left( t\right) }{%
\left( \frac{b}{\Gamma \left( \alpha \right) }+t\right) ^{1-\alpha }}dt. 
\tag{8}  \label{a12}
\end{equation}

\bigskip

\noindent \textbf{Theorem 4.} \cite{gomez} Let $\func{Re}\left( \beta
\right) \geq 0,$ $n=\left[ \func{Re}\left( \beta \right) \right] +1,$ $f\in
C_{\alpha ,a}^{n}\left( \left[ a,b\right] \right) .$ Then a different type
of Riemann-Liouville fractional conformable derivatives are defined as
follows,%
\begin{equation}
_{a}^{AR\beta }D^{\alpha }f\left( x\right) =\frac{_{\quad a}^{A\quad
n}D^{\alpha }}{\Gamma \left( n-\beta \right) }\dint \limits_{a}^{x}\left( 
\frac{\left( x+\frac{a}{\Gamma \left( \alpha \right) }\right) ^{\alpha
}-\left( t+\frac{a}{\Gamma \left( \alpha \right) }\right) ^{\alpha }}{\alpha 
}\right) ^{n-\beta -1}\frac{f\left( t\right) }{\left( t+\frac{a}{\Gamma
\left( \alpha \right) }\right) ^{1-\alpha }}dt,  \tag{9}  \label{a13}
\end{equation}%
and 
\begin{equation}
^{AR\beta }D_{b}^{\alpha }f\left( x\right) =\frac{\left( -1\right) _{\quad
\quad }^{nA\quad n}D_{b}^{\alpha }}{\Gamma \left( n-\beta \right) }\dint
\limits_{x}^{b}\left( \frac{\left( \frac{b}{\Gamma \left( \alpha \right) }%
+t\right) ^{\alpha }-\left( \frac{b}{\Gamma \left( \alpha \right) }+x\right)
^{\alpha }}{\alpha }\right) ^{n-\beta -1}\frac{f\left( t\right) }{\left( 
\frac{b}{\Gamma \left( \alpha \right) }+t\right) ^{1-\alpha }}dt.  \tag{10}
\label{a14}
\end{equation}

\bigskip

\noindent \textbf{Theorem 5.} \cite{gomez} Let $f\in C_{\alpha ,a}^{n}\left( %
\left[ a,b\right] \right) ,$ $\beta \in 
\mathbb{C}
.$ Then the following property is valid, 
\begin{equation*}
_{a}^{\beta }I^{\alpha }\left( _{\quad a}^{AC\beta }D^{\alpha }f\left(
t\right) \right) =f\left( t\right) -\sum_{k=0}^{n-1}\frac{%
_{a}^{k}D_{t}^{\alpha }f\left( a\right) \left( t-a\right) ^{\alpha k}}{%
\alpha ^{k}k!},
\end{equation*}%
and 
\begin{equation*}
_{a}^{\beta }I^{\alpha }\left( _{a}^{AC\beta }D^{\alpha }f\left( t\right)
\right) =f\left( t\right) -\sum_{k=0}^{n-1}\frac{\left( -1\right) ^{k}\quad
_{t}^{k}D_{b}^{\alpha }f\left( b\right) \left( b-t\right) ^{\alpha k}}{%
\alpha ^{k}k!}.
\end{equation*}

\section{Main Results}

In this section, we find exact analytical solutions of Newton's law of
cooling with Caputo fractional conformable and the different type of
conformable derivative.

\subsection{\textbf{Newton's Law of Cooling with Newly Defined Fractional
Conformable Derivative}}

Let's consider Caputo fractional conformable derivative defined in $\left( %
\ref{a7}\right) .$ So, we obtain exact analytical solution of the Newton's
law of cooling equation. Let's consider the initial value problem%
\begin{eqnarray}
_{a}^{C\beta }D_{t}^{\alpha }T\left( t\right) &=&-k\left( T-T_{m}\right) , 
\TCItag{11}  \label{b1} \\
T\left( a\right) &=&T_{0}.  \TCItag{12}  \label{b2}
\end{eqnarray}

\bigskip

\textbf{Solution.} Using the inverse operator of $_{a}^{C\beta
}D_{t}^{\alpha }$ to equation $\left( \ref{b1}\right) ,$ we get 
\begin{equation*}
_{a}^{\beta }I^{\alpha C\beta }D_{t}^{\alpha }T\left( t\right) =_{a}^{\beta
}I_{t}^{\alpha }\left \{ -k\left( T-T_{m}\right) \right \} .
\end{equation*}%
Considering the Theorem 2 and the initial condition $\left( \ref{b2}\right)
, $ we have

\begin{equation*}
T\left( t\right) =T\left( a\right) -_{a}^{\beta }I_{t}^{\alpha }\left \{
k\left( T-T_{m}\right) \right \} .
\end{equation*}%
Then%
\begin{equation*}
T_{n+1}\left( t\right) =T_{0}-k_{0}^{\beta }I_{t}^{\alpha }\left \{ \left(
T_{n}-T_{m}\right) \right \} ,\text{ \ }n=0,1,2,..
\end{equation*}%
For $n=0,$ we can write%
\begin{equation}
T_{1}\left( t\right) =T_{0}-k_{a}^{\beta }I_{t}^{\alpha }\left \{ \left(
T_{0}-T_{m}\right) \right \}  \tag{13}  \label{b3}
\end{equation}%
where%
\begin{equation*}
_{a}^{\beta }I_{t}^{\alpha }\left \{ \left( T_{0}-T_{m}\right) \right \} =%
\frac{\left( T_{0}-T_{m}\right) }{\Gamma \left( \beta \right) }\dint
\limits_{a}^{t}\left( \frac{\left( t-a\right) ^{\alpha }-\left( x-a\right)
^{\alpha }}{\alpha }\right) ^{\beta -1}\frac{dx}{\left( x-a\right)
^{1-\alpha }}.
\end{equation*}%
Applying the change of variable $u=\left( \frac{x-a}{t-a}\right) ^{\alpha },$
we have%
\begin{equation}
I_{t}^{\alpha }\left \{ \left( T_{0}-T_{m}\right) \right \} =\frac{\left(
T_{0}-T_{m}\right) \left( t-a\right) ^{\alpha \beta }}{\alpha ^{\beta
}\Gamma \left( \beta +1\right) }  \tag{14}  \label{b4}
\end{equation}%
Substituting equation $\left( \ref{b4}\right) $ into $\left( \ref{b3}\right)
,$we have%
\begin{equation*}
T_{1}\left( t\right) =T_{0}-\frac{k\left( T_{0}-T_{m}\right) \left(
t-a\right) ^{\alpha \beta }}{\alpha ^{\beta }\Gamma \left( \beta +1\right) }.
\end{equation*}%
For $n=1,$ we get%
\begin{eqnarray}
T_{2}\left( t\right) &=&T_{0}-k_{a}^{\beta }I_{t}^{\alpha }\left \{ \left(
T_{1}-T_{m}\right) \right \}  \notag \\
&=&T_{0}-k_{a}^{\beta }I_{t}^{\alpha }\left \{ \left( T_{0}-T_{m}-\frac{%
k\left( T_{0}-T_{m}\right) \left( t-a\right) ^{\alpha \beta }}{\alpha
^{\beta }\Gamma \left( \beta +1\right) }\right) \right \}  \notag \\
&=&T_{0}-k_{a}^{\beta }I_{t}^{\alpha }\left \{ T_{0}-T_{m}\right \} +\frac{%
k^{2}\left( T_{0}-T_{m}\right) }{\alpha ^{\beta }\Gamma \left( \beta
+1\right) }_{a}^{\beta }I_{t}^{\alpha }\left \{ \left( t-a\right) ^{\alpha
\beta }\right \} ,  \TCItag{15}  \label{b5}
\end{eqnarray}%
where $_{a}^{\beta }I_{t}^{\alpha }\left \{ \left( t-a\right) ^{\alpha \beta
}\right \} =\frac{\Gamma \left( \beta +1\right) \left( t-a\right) ^{2\alpha
\beta }}{\alpha ^{\beta }\Gamma \left( 2\beta +1\right) },$ then we can
rewrite equation $\left( \ref{b5}\right) $ 
\begin{eqnarray*}
&=&T_{0}-\frac{k\left( T_{0}-T_{m}\right) \left( t-a\right) ^{\alpha \beta }%
}{\alpha ^{\beta }\Gamma \left( \beta +1\right) }+\frac{k^{2}\left(
T_{0}-T_{m}\right) \left( t-a\right) ^{2\alpha \beta }}{\alpha ^{2\beta
}\Gamma \left( 2\beta +1\right) } \\
&=&T_{0}\left( 1-\frac{k\left( t-a\right) ^{\alpha \beta }}{\alpha ^{\beta
}\Gamma \left( \beta +1\right) }+\frac{k^{2}\left( t-a\right) ^{2\alpha
\beta }}{\alpha ^{2\beta }\Gamma \left( 2\beta +1\right) }\right) + \\
&&+\frac{T_{m}k\left( t-a\right) ^{\alpha \beta }}{\alpha ^{\beta }}\left( 
\frac{1}{\beta \Gamma \left( \beta \right) }-\frac{k\left( t-a\right)
^{\alpha \beta }}{\alpha ^{\beta }2\beta \Gamma \left( 2\beta \right) }%
\right) .
\end{eqnarray*}%
Proceeding inductively we have 
\begin{eqnarray*}
T_{n}\left( t\right) &=&T_{0}\left( 1-\frac{k\left( t-a\right) ^{\alpha
\beta }}{\alpha ^{\beta }\Gamma \left( \beta +1\right) }+\frac{k^{2}\left(
t-a\right) ^{2\alpha \beta }}{\alpha ^{2\beta }\Gamma \left( 2\beta
+1\right) }-...\right) \\
&&+\frac{T_{m}k\left( t-a\right) ^{\alpha \beta }}{\alpha ^{\beta }}\left( 
\frac{1}{\beta \Gamma \left( \beta \right) }-\frac{k\left( t-a\right)
^{\alpha \beta }}{\alpha ^{\beta }2\beta \Gamma \left( 2\beta \right) }%
+...\right) \\
&=&T_{0}\dsum \limits_{z=0}^{n}\frac{\left( -1\right) ^{z}k^{z}\left(
t-a\right) ^{z\alpha \beta }}{\alpha ^{z\beta }\Gamma \left( z\beta
+1\right) }+\frac{kT_{m}\left( t-a\right) ^{\alpha \beta }}{\alpha ^{\beta }}%
\dsum \limits_{z=0}^{n}\frac{\left( -1\right) ^{z}k^{z}\left( t-a\right)
^{z\alpha \beta }}{\alpha ^{z\beta }\left( z+1\right) \beta \Gamma \left(
z\beta +\beta \right) }.
\end{eqnarray*}%
Therefore, as $n\rightarrow \infty $, we find%
\begin{equation*}
T\left( t\right) =T_{0}E_{\beta ,1}\left( -\frac{k}{\alpha ^{\beta }}\left(
t-a\right) ^{\alpha \beta }\right) +\frac{kT_{m}\left( t-a\right) ^{\alpha
\beta }}{\alpha ^{\beta }}\dsum \limits_{z=0}^{\infty }\frac{\left(
-1\right) ^{z}k^{z}\left( t-a\right) ^{z\alpha \beta }}{\alpha ^{z\beta
}\left( z+1\right) \beta \Gamma \left( z\beta +\beta \right) },
\end{equation*}%
where $E_{\beta ,1}\left( t\right) $ is Mittag-Leffler function \cite{podlub}%
.

\bigskip

Now, let's consider the different type of Caputo fractional conformable
derivatives defined in $\left( \ref{a11}\right) .$ We obtain the analytical
solution of the Newton's law of cooling. Considering the initial value
problem%
\begin{eqnarray*}
_{a}^{AC\beta }D_{t}^{\alpha }T\left( t\right) &=&-k\left( T-T_{m}\right) ,
\\
T\left( a\right) &=&T_{0}.
\end{eqnarray*}%
\bigskip

\textbf{Solution. }If we apply similar arguments used in the proof of
problem $\left( \ref{b1}-\ref{b2}\right) $, then we have%
\begin{equation*}
T\left( t\right) =T_{0}E_{\beta ,1}\left( -\frac{k}{\alpha ^{\beta }}\left(
t+\frac{a}{\Gamma \left( \alpha \right) }\right) ^{\alpha \beta }\right) +%
\frac{kT_{m}\left( t+\frac{a}{\Gamma \left( \alpha \right) }\right) ^{\alpha
\beta }}{\alpha ^{\beta }}\dsum \limits_{z=0}^{\infty }\frac{\left(
-1\right) ^{z}k^{z}\left( t+\frac{a}{\Gamma \left( \alpha \right) }\right)
^{z\alpha \beta }}{\alpha ^{z\beta }\left( z+1\right) \beta \Gamma \left(
z\beta +\beta \right) }.
\end{equation*}

\subsection{\textbf{Comparative Analysis and Discussions}}

In this section, we use experimental data tested by Gieseking \cite{eliza}.
Gieseking \cite{eliza} used three beakers of water in volume of $100$, $300$
and $800$ $ml$ and measured temperature every minute for $35$ minutes in a
constant ambient temperature $23^{o}C.$ We study to find the optimal
fractional orders which fit better with real data. Results show that
Newton's law of cooling with fractional conformable derivative gives better
results to integer order derivative. Results are given comparatively to
Newton's law of cooling with integer order and experimental data and also,
fractional conformable derivative's advantages are supported by numerical
simulations and error analysis. We try different fractional orders $\left(
\alpha =0.9,\beta =0.9\right) ,\left( \alpha =0.9,\beta =0.95\right) ,$ $%
\left( \alpha =0.91,\beta =0.91\right) $ and $\left( \alpha =0.92,\beta
=0.92\right) $ for finding optimal order to fit real data.

Let's consider Newton's law of cooling with integer order derivative and its
solution is as follows%
\begin{eqnarray*}
T\left( t\right) &=&T_{m}+\left( T_{0}-T_{m}\right) e^{\widetilde{k}t}, \\
T_{0} &=&100^{o}C, \\
T_{m} &=&23^{o}C, \\
\widetilde{k} &=&\frac{\ln \left( \frac{T\left( t\right) -T_{m}}{T_{0}-T_{m}}%
\right) }{t}.
\end{eqnarray*}%
The convection coefficient $\widetilde{k}$ for Newton's law of cooling with
integer order can be found analytically, but the convection coefficient $k$
for Newton's law of cooling with fractional order can be found
approximately. The convection coefficient $\widetilde{k}$ for three beakers
of water in volume of $100,300$ and $800$ $ml$ is found as respectively $%
\widetilde{k}=0.0676$, $\widetilde{k}=0.0447$ and $\widetilde{k}=0.0327.$

Assume that temperatures of these three beakers of water in volume of $%
100,300$ and $800$ $ml$ are known as $45^{o}C$, $55^{o}C$ and $63^{o}C$
after $20$ minutes for finding the approximate value of convection
coefficient $k,$ and $k$ will change for each value $\alpha $ and order $%
\beta $.

If we consider beaker of water in volume of $100$ $ml$, then we observe that
the optimal fractional order $\alpha =0.9,\beta =0.9$ to fit real data with
error analysis and simulation in $Fig.1,$ $Table1-2-3-4.$

If we consider beaker of water in volume of $300$ $ml$, then we observe that
the optimal fractional order $\alpha =0.9,\beta =0.9$ to fit real data with
error analysis and simulation in $Fig.2,$ $Table5-6-7-8.$

If we consider beaker of water in volume of $800$ $ml$, then we observe that
the optimal fractional order $\alpha =0.9,\beta =0.95$ to fit real data with
error analysis and simulation in $Fig.3,$ $Table9-10-11-12.$

Finally, we compare these results similarly under any value $\alpha $ and
any order $\beta $ in $Fig.4,$ $Fig.5$ and $Fig.6.$

\noindent \textbf{.}$\underset{Fig.1.:\text{Comparison of cooling water }%
V=100ml}{%
\begin{array}{cc}
\FRAME{itbpF}{2.8454in}{2.2157in}{0in}{}{}{Figure}{\special{language
"Scientific Word";type "GRAPHIC";maintain-aspect-ratio TRUE;display
"USEDEF";valid_file "T";width 2.8454in;height 2.2157in;depth
0in;original-width 6.052in;original-height 4.708in;cropleft "0";croptop
"1";cropright "1";cropbottom "0";tempfilename
'PAXCU606.wmf';tempfile-properties "PR";}} & \FRAME{itbpF}{2.794in}{2.2157in%
}{0in}{}{}{Figure}{\special{language "Scientific Word";type
"GRAPHIC";maintain-aspect-ratio TRUE;display "USEDEF";valid_file "T";width
2.794in;height 2.2157in;depth 0in;original-width 5.9689in;original-height
4.7288in;cropleft "0";croptop "1";cropright "1";cropbottom "0";tempfilename
'PAXCY107.wmf';tempfile-properties "PR";}} \\ 
\FRAME{itbpF}{2.7984in}{2.2192in}{0in}{}{}{Figure}{\special{language
"Scientific Word";type "GRAPHIC";maintain-aspect-ratio TRUE;display
"USEDEF";valid_file "T";width 2.7984in;height 2.2192in;depth
0in;original-width 6.0727in;original-height 4.8127in;cropleft "0";croptop
"1";cropright "1";cropbottom "0";tempfilename
'PAXHKM02.wmf';tempfile-properties "PR";}} & \FRAME{itbpF}{2.9332in}{2.26in}{%
0in}{}{}{Figure}{\special{language "Scientific Word";type
"GRAPHIC";maintain-aspect-ratio TRUE;display "USEDEF";valid_file "T";width
2.9332in;height 2.26in;depth 0in;original-width 6.1358in;original-height
4.7184in;cropleft "0";croptop "1";cropright "1";cropbottom "0";tempfilename
'PAXHLF03.wmf';tempfile-properties "PR";}}%
\end{array}%
}$\newline
$\underset{Table1.V=100ml,\alpha =0.92,\beta =0.92,k\simeq 0.0951}{%
\begin{tabular}{|c|c|c|c|c|c|c|c|c|c|}
\hline
Time(min.) & $0$ & $5$ & $10$ & $15$ & $20$ & $25$ & $30$ & $35$ & SSE \\ 
\hline
Experimental($^{o}C)$ & $100$ & $70$ & $57$ & $50$ & $45$ & $41$ & $38$ & $%
36 $ &  \\ \hline
Classical($^{o}C)$ & $100$ & $77.916$ & $62.1659$ & $50.9329$ & $42.9216$ & $%
37.208$ & $33.1331$ & $30.2269$ &  \\ \hline
Fractional$(^{o}C)$ & $100$ & $74.2755$ & $60.4967$ & $51.379$ & $45.$ & $%
40.3893$ & $36.9786$ & $34.4092$ &  \\ \hline
Error(clsc-expr) & $0$ & $0.1015$ & $0.0830$ & $0.0183$ & $0.04842$ & $%
0.1019 $ & $0.1468$ & $0.1909$ & $4.6093$ \\ \hline
Error(frac-expr)) & $0$ & $0.0575$ & $0.0578$ & $0.0268$ & $0$ & $0.0151$ & $%
0.0276$ & $0.0462$ & $1.0098$ \\ \hline
\end{tabular}%
}$

SSE:Error Sum of Squares\newline

\noindent $\underset{Table2.V=100ml,\alpha =0.91,\beta =0.91,k\simeq 0.1000}{%
\begin{tabular}{|c|c|c|c|c|c|c|c|c|c|}
\hline
Time(min.) & $0$ & $5$ & $10$ & $15$ & $20$ & $25$ & $30$ & $35$ & SSE \\ 
\hline
Experimental($^{o}C)$ & $100$ & $70$ & $57$ & $50$ & $45$ & $41$ & $38$ & $%
36 $ &  \\ \hline
Classical($^{o}C)$ & $100$ & $77.916$ & $62.1659$ & $50.9329$ & $42.9216$ & $%
37.208$ & $33.1331$ & $30.2269$ &  \\ \hline
Fractional$(^{o}C)$ & $100$ & $73.6236$ & $60.0584$ & $51.1824$ & $45.$ & $%
40.5348$ & $37.2262$ & $34.7252$ &  \\ \hline
Error(clsc-expr) & $0$ & $0.1015$ & $0.0830$ & $0.0183$ & $0.04842$ & $%
0.1019 $ & $0.1468$ & $0.1909$ & $4.6093$ \\ \hline
Error(frac-expr)) & $0$ & $0.04921$ & $0.0509$ & $0.0231$ & $0$ & $0.0114$ & 
$0.0207$ & $0.0367$ & $0.7311$ \\ \hline
\end{tabular}%
}$\newline

\bigskip

\noindent $\underset{Table3.V=100ml,\alpha =0.9,\beta =0.95,k\simeq 0.0892}{%
\begin{tabular}{|c|c|c|c|c|c|c|c|c|c|}
\hline
Time(min.) & $0$ & $5$ & $10$ & $15$ & $20$ & $25$ & $30$ & $35$ & SSE \\ 
\hline
Experimental($^{o}C)$ & $100$ & $70$ & $57$ & $50$ & $45$ & $41$ & $38$ & $%
36 $ &  \\ \hline
Classical($^{o}C)$ & $100$ & $77.916$ & $62.1659$ & $50.9329$ & $42.9216$ & $%
37.208$ & $33.1331$ & $30.2269$ &  \\ \hline
Fractional$(^{o}C)$ & $100$ & $74.8976$ & $60.9808$ & $51.6192$ & $45.$ & $%
40.1844$ & $36.6107$ & $33.9174$ &  \\ \hline
Error(clsc-expr) & $0$ & $0.1015$ & $0.0830$ & $0.0183$ & $0.04842$ & $%
0.1019 $ & $0.1468$ & $0.1909$ & $4.6093$ \\ \hline
Error(frac-expr)) & $0$ & $0.0653$ & $0.0652$ & $0.0313$ & $0$ & $0.0202$ & $%
0.0379$ & $0.0614$ & $1.3718$ \\ \hline
\end{tabular}%
}$\newline

\bigskip

\noindent $\underset{Table4.V=100ml,\alpha =0.9,\beta =0.9,k\simeq 0.1052}{%
\begin{tabular}{|c|c|c|c|c|c|c|c|c|c|}
\hline
Time(min.) & $0$ & $5$ & $10$ & $15$ & $20$ & $25$ & $30$ & $35$ & SSE \\ 
\hline
Experimental($^{o}C)$ & $100$ & $70$ & $57$ & $50$ & $45$ & $41$ & $38$ & $%
36 $ &  \\ \hline
Classical($^{o}C)$ & $100$ & $77.916$ & $62.1659$ & $50.9329$ & $42.9216$ & $%
37.208$ & $33.1331$ & $30.2269$ &  \\ \hline
Fractional$(^{o}C)$ & $100$ & $72.9684$ & $59.6254$ & $50.9897$ & $45.$ & $%
40.6766$ & $37.4676$ & $35.0336$ &  \\ \hline
Error(clsc-expr) & $0$ & $0.1015$ & $0.0830$ & $0.0183$ & $0.04842$ & $%
0.1019 $ & $0.1468$ & $0.1909$ & $4.6093$ \\ \hline
Error(frac-expr)) & $0$ & $0.04068$ & $0.0440$ & $0.0194$ & $0$ & $0.0079$ & 
$0.0142$ & $0.0275$ & $0.5001$ \\ \hline
\end{tabular}%
}$\newline
$\underset{Fig.2.:\text{Comparison of cooling water }V=300ml}{%
\begin{array}{cc}
\FRAME{itbpF}{2.5563in}{2.009in}{0in}{}{}{Figure}{\special{language
"Scientific Word";type "GRAPHIC";maintain-aspect-ratio TRUE;display
"USEDEF";valid_file "T";width 2.5563in;height 2.009in;depth
0in;original-width 5.9689in;original-height 4.6873in;cropleft "0";croptop
"1";cropright "1";cropbottom "0";tempfilename
'PAXDLV0E.wmf';tempfile-properties "PR";}} & \FRAME{itbpF}{2.5793in}{2.025in%
}{0in}{}{}{Figure}{\special{language "Scientific Word";type
"GRAPHIC";maintain-aspect-ratio TRUE;display "USEDEF";valid_file "T";width
2.5793in;height 2.025in;depth 0in;original-width 6.0312in;original-height
4.7288in;cropleft "0";croptop "1";cropright "1";cropbottom "0";tempfilename
'PAXHF100.wmf';tempfile-properties "PR";}} \\ 
\FRAME{itbpF}{2.7647in}{2.1376in}{0in}{}{}{Figure}{\special{language
"Scientific Word";type "GRAPHIC";maintain-aspect-ratio TRUE;display
"USEDEF";valid_file "T";width 2.7647in;height 2.1376in;depth
0in;original-width 6.1246in;original-height 4.7288in;cropleft "0";croptop
"1";cropright "1";cropbottom "0";tempfilename
'PAXDS40H.wmf';tempfile-properties "PR";}} & \FRAME{itbpF}{2.7656in}{2.1252in%
}{0in}{}{}{Figure}{\special{language "Scientific Word";type
"GRAPHIC";maintain-aspect-ratio TRUE;display "USEDEF";valid_file "T";width
2.7656in;height 2.1252in;depth 0in;original-width 6.1358in;original-height
4.708in;cropleft "0";croptop "1";cropright "1";cropbottom "0";tempfilename
'PAXDTU0I.wmf';tempfile-properties "PR";}}%
\end{array}%
}$\newline

\noindent $\underset{Table5.V=300ml,\alpha =0.92,\beta =0.92,k\simeq 0.0651}{%
\begin{tabular}{|c|c|c|c|c|c|c|c|c|c|}
\hline
Time(min.) & $0$ & $5$ & $10$ & $15$ & $20$ & $25$ & $30$ & $35$ & SSE \\ 
\hline
Experimental($^{o}C)$ & $100$ & $81$ & $70$ & $61$ & $55$ & $51$ & $47$ & $%
45 $ &  \\ \hline
Classical($^{o}C)$ & $100$ & $84.578$ & $72.2449$ & $62.3819$ & $54.4943$ & $%
48.1864$ & $43.142$ & $39.1078$ &  \\ \hline
Fractional$(^{o}C)$ & $100$ & $81.1929$ & $69.7805$ & $61.4169$ & $55.$ & $%
49.9507$ & $45.9091$ & $42.6325$ &  \\ \hline
Error(clsc-expr) & $0$ & $0.0423$ & $0.0310$ & $0.0221$ & $0.0092$ & $0.0583$
& $0.0894$ & $0.1506$ & $2.1534$ \\ \hline
Error(frac-expr)) & $0$ & $0.0023$ & $0.0031$ & $0.0067$ & $0$ & $0.0210$ & $%
0.0237$ & $0.0555$ & $0.2265$ \\ \hline
\end{tabular}%
}$\newline

$\bigskip $

\noindent $\underset{Table6.V=300ml,\alpha =0.91,\beta =0.91,k\simeq 0.06830}%
{%
\begin{tabular}{|c|c|c|c|c|c|c|c|c|c|}
\hline
Time(min.) & $0$ & $5$ & $10$ & $15$ & $20$ & $25$ & $30$ & $35$ & SSE \\ 
\hline
Experimental($^{o}C)$ & $100$ & $81$ & $70$ & $61$ & $55$ & $51$ & $47$ & $%
45 $ &  \\ \hline
Classical($^{o}C)$ & $100$ & $84.578$ & $72.2449$ & $62.3819$ & $54.4943$ & $%
48.1864$ & $43.142$ & $39.1078$ &  \\ \hline
Fractional$(^{o}C)$ & $100$ & $80.7165$ & $69.4317$ & $61.2478$ & $55$ & $%
50.0939$ & $46.1678$ & $42.9812$ &  \\ \hline
Error(clsc-expr) & $0$ & $0.0423$ & $0.0310$ & $0.0221$ & $0.0092$ & $0.0583$
& $0.0894$ & $0.1506$ & $2.1534$ \\ \hline
Error(frac-expr)) & $0$ & $0.0035$ & $0.0081$ & $0.0040$ & $0$ & $0.0180$ & $%
0.0180$ & $0.0469$ & $0.1681$ \\ \hline
\end{tabular}%
}$\newline

$\bigskip $

\noindent $\underset{Table7.V=300ml,\alpha =0.9,\beta =0.95,k\simeq 0.0616}{%
\begin{tabular}{|c|c|c|c|c|c|c|c|c|c|}
\hline
Time(min.) & $0$ & $5$ & $10$ & $15$ & $20$ & $25$ & $30$ & $35$ & SSE \\ 
\hline
Experimental($^{o}C)$ & $100$ & $81$ & $70$ & $61$ & $55$ & $51$ & $47$ & $%
45 $ &  \\ \hline
Classical($^{o}C)$ & $100$ & $84.578$ & $72.2449$ & $62.3819$ & $54.4943$ & $%
48.1864$ & $43.142$ & $39.1078$ &  \\ \hline
Fractional$(^{o}C)$ & $100$ & $81.5723$ & $70.0951$ & $61.5832$ & $55$ & $%
49.7909$ & $45.6057$ & $42.205$ &  \\ \hline
Error(clsc-expr) & $0$ & $0.0423$ & $0.0310$ & $0.0221$ & $0.0092$ & $0.0583$
& $0.0894$ & $0.1506$ & $2.1534$ \\ \hline
Error(frac-expr)) & $0$ & $0.0070$ & $0.0013$ & $0.0040$ & $0$ & $0.0242$ & $%
0.0305$ & $0.0662$ & $0.3304$ \\ \hline
\end{tabular}%
}$\newline

$\bigskip $

\noindent $\underset{Table8.V=300ml,\alpha =0.9,\beta =0.9,k\simeq 0.07160}{%
\begin{tabular}{|c|c|c|c|c|c|c|c|c|c|}
\hline
Time(min.) & $0$ & $5$ & $10$ & $15$ & $20$ & $25$ & $30$ & $35$ & SSE \\ 
\hline
Experimental($^{o}C)$ & $100$ & $81$ & $70$ & $61$ & $55$ & $51$ & $47$ & $%
45 $ &  \\ \hline
Classical($^{o}C)$ & $100$ & $84.578$ & $72.2449$ & $62.3819$ & $54.4943$ & $%
48.1864$ & $43.142$ & $39.1078$ &  \\ \hline
Fractional$(^{o}C)$ & $100$ & $80.2362$ & $69.0853$ & $61.0812$ & $55$ & $%
50.2341$ & $46.4208$ & $43.3225$ &  \\ \hline
Error(clsc-expr) & $0$ & $0.0423$ & $0.0310$ & $0.0221$ & $0.0092$ & $0.0583$
& $0.0894$ & $0.1506$ & $2.1534$ \\ \hline
Error(frac-expr)) & $0$ & $0.0095$ & $0.0132$ & $0.0013$ & $0$ & $0.0152$ & $%
0.0124$ & $0.0387$ & $0.1434$ \\ \hline
\end{tabular}%
}$\newline

\noindent $\underset{Fig.3.:\text{Comparison of cooling water }V=800ml}{%
\begin{array}{cc}
\FRAME{itbpF}{2.8685in}{2.2175in}{0in}{}{}{Figure}{\special{language
"Scientific Word";type "GRAPHIC";maintain-aspect-ratio TRUE;display
"USEDEF";valid_file "T";width 2.8685in;height 2.2175in;depth
0in;original-width 4.6043in;original-height 3.5518in;cropleft "0";croptop
"1";cropright "1";cropbottom "0";tempfilename
'PAXE2Z0K.wmf';tempfile-properties "PR";}} & \FRAME{itbpF}{2.8029in}{2.1988in%
}{0in}{}{}{Figure}{\special{language "Scientific Word";type
"GRAPHIC";maintain-aspect-ratio TRUE;display "USEDEF";valid_file "T";width
2.8029in;height 2.1988in;depth 0in;original-width 4.5731in;original-height
3.5829in;cropleft "0";croptop "1";cropright "1";cropbottom "0";tempfilename
'PAXE480L.wmf';tempfile-properties "PR";}} \\ 
\FRAME{itbpF}{2.9341in}{2.2982in}{0in}{}{}{Figure}{\special{language
"Scientific Word";type "GRAPHIC";maintain-aspect-ratio TRUE;display
"USEDEF";valid_file "T";width 2.9341in;height 2.2982in;depth
0in;original-width 4.6354in;original-height 3.6253in;cropleft "0";croptop
"1";cropright "1";cropbottom "0";tempfilename
'PAXE4X0M.wmf';tempfile-properties "PR";}} & \FRAME{itbpF}{2.8401in}{2.2964in%
}{0in}{}{}{Figure}{\special{language "Scientific Word";type
"GRAPHIC";maintain-aspect-ratio TRUE;display "USEDEF";valid_file "T";width
2.8401in;height 2.2964in;depth 0in;original-width 4.6561in;original-height
3.7602in;cropleft "0";croptop "1";cropright "1";cropbottom "0";tempfilename
'PAXE5H0N.wmf';tempfile-properties "PR";}}%
\end{array}%
}$\newline

\noindent $\underset{Table9.V=800ml,\alpha =0.92,\beta =0.92,k\simeq 0.0.0479%
}{%
\begin{tabular}{|c|c|c|c|c|c|c|c|c|c|}
\hline
Time(min.) & $0$ & $5$ & $10$ & $15$ & $20$ & $25$ & $30$ & $35$ & SSE \\ 
\hline
Experimental($^{o}C)$ & $100$ & $88$ & $76$ & $70$ & $63$ & $59$ & $54$ & $%
50 $ &  \\ \hline
Classical($^{o}C)$ & $100$ & $88.3858$ & $78.5234$ & $70.1487$ & $63.0371$ & 
$56.9981$ & $51.8701$ & $47.5155$ &  \\ \hline
Fractional$(^{o}C)$ & $100$ & $85.602$ & $76.2156$ & $68.9149$ & $63$ & $%
58.1028$ & $53.9902$ & $50.5007$ &  \\ \hline
Error(clsc-expr) & $0$ & $0.0043$ & $0.0321$ & $0.0021$ & $0.0005$ & $0.0351$
& $0.0410$ & $0.0522$ & $0.5904$ \\ \hline
Error(frac-expr)) & $0$ & $0.02801$ & $0.0028$ & $0.0157$ & $0$ & $0.0154$ & 
$0.0001$ & $0.0099$ & $0.2230$ \\ \hline
\end{tabular}%
}$\newline

$\bigskip $

\noindent $\underset{Table10.V=800ml,\alpha =0.91,\beta =0.91,k\simeq 0.0502}%
{%
\begin{tabular}{|c|c|c|c|c|c|c|c|c|c|}
\hline
Time(min.) & $0$ & $5$ & $10$ & $15$ & $20$ & $25$ & $30$ & $35$ & SSE \\ 
\hline
Experimental($^{o}C)$ & $100$ & $88$ & $76$ & $70$ & $63$ & $59$ & $54$ & $%
50 $ &  \\ \hline
Classical($^{o}C)$ & $100$ & $88.3858$ & $78.5234$ & $70.1487$ & $63.0371$ & 
$56.9981$ & $51.8701$ & $47.5155$ &  \\ \hline
Fractional$(^{o}C)$ & $100$ & $85.2367$ & $75.9343$ & $68.7721$ & $63$ & $%
58.2343$ & $54.2368$ & $50.8452$ &  \\ \hline
Error(clsc-expr) & $0$ & $0.0043$ & $0.0321$ & $0.0021$ & $0.0005$ & $0.0351$
& $0.0410$ & $0.0522$ & $0.5904$ \\ \hline
Error(frac-expr)) & $0$ & $0.0324$ & $0.0008$ & $0.0178$ & $0.$ & $0.0131$ & 
$0.0043$ & $0.0166$ & $0.2917$ \\ \hline
\end{tabular}%
}$\newline

$\bigskip $

\noindent $\underset{Table11.V=800ml,\alpha =0.9,\beta =0.95,k\simeq 0.0456}{%
\begin{tabular}{|c|c|c|c|c|c|c|c|c|c|}
\hline
Time(min.) & $0$ & $5$ & $10$ & $15$ & $20$ & $25$ & $30$ & $35$ & SSE \\ 
\hline
Experimental($^{o}C)$ & $100$ & $88$ & $76$ & $70$ & $63$ & $59$ & $54$ & $%
50 $ &  \\ \hline
Classical($^{o}C)$ & $100$ & $88.3858$ & $78.5234$ & $70.1487$ & $63.0371$ & 
$56.9981$ & $51.8701$ & $47.5155$ &  \\ \hline
Fractional$(^{o}C)$ & $100$ & $85.8602$ & $76.4366$ & $69.0358$ & $63$ & $%
57.9785$ & $53.7464$ & $50.1462$ &  \\ \hline
Error(clsc-expr) & $0$ & $0.0043$ & $0.0321$ & $0.0021$ & $0.0005$ & $0.0351$
& $0.0410$ & $0.0522$ & $0.5904$ \\ \hline
Error(frac-expr)) & $0$ & $0.0249$ & $0.0057$ & $0.013$ & $0.$ & $0.0176$ & $%
0.0047$ & $0.0029$ & $0.1896$ \\ \hline
\end{tabular}%
}$\newline

$\bigskip $

\noindent $\underset{Table12.V=800ml,;\alpha =0.9,\beta =0.9,k\simeq 0.0525}{%
\begin{tabular}{|r|r|r|r|r|r|r|r|r|r|}
\hline
Time(min.) & $0$ & $5$ & $10$ & $15$ & $20$ & $25$ & $30$ & $35$ & SSE \\ 
\hline
Experimental($^{o}C)$ & $100$ & $88$ & $76$ & $70$ & $63$ & $59$ & $54$ & $%
50 $ &  \\ \hline
Classical$(^{o}C)$ & $100$ & $88.3858$ & $78.5234$ & $70.1487$ & $63.0371$ & 
$56.9981$ & $51.8701$ & $47.5155$ &  \\ \hline
Fractional$(^{o}C)$ & $100$ & $84.8678$ & $75.6542$ & $68.631$ & $63$ & $%
58.3632$ & $54.4785$ & $51.1827$ &  \\ \hline
Error(clsc-expr) & $0$ & $0.0043$ & $0.0321$ & $0.0021$ & $0.0005$ & $0.0351$
& $0.0410$ & $0.0522$ & $0.5904$ \\ \hline
Error(frac-expr)) & $0$ & $0.0369$ & $0.0045$ & $0.0199$ & $0.$ & $0.0109$ & 
$0.0087$ & $0.0231$ & $0.3843$ \\ \hline
\end{tabular}%
}$\newline
\newline
\noindent $%
\begin{array}{ccc}
\underset{Fig.4.V=100ml}{\FRAME{itbpF}{2.0099in}{1.3793in}{0pt}{}{}{Figure}{%
\special{language "Scientific Word";type "GRAPHIC";maintain-aspect-ratio
TRUE;display "USEDEF";valid_file "T";width 2.0099in;height 1.3793in;depth
0pt;original-width 6.979in;original-height 4.7816in;cropleft "0";croptop
"1";cropright "1";cropbottom "0";tempfilename
'PAXH7V0U.wmf';tempfile-properties "PR";}}} & \underset{Fig.5.V=300ml}{%
\FRAME{itbpF}{2.0445in}{1.3926in}{0pt}{}{}{Figure}{\special{language
"Scientific Word";type "GRAPHIC";maintain-aspect-ratio TRUE;display
"USEDEF";valid_file "T";width 2.0445in;height 1.3926in;depth
0pt;original-width 6.9272in;original-height 4.708in;cropleft "0";croptop
"1";cropright "1";cropbottom "0";tempfilename
'PAXH7V0V.wmf';tempfile-properties "PR";}}} & \underset{Fig.6.V=800ml}{%
\FRAME{itbpF}{2.1607in}{1.4502in}{0.0346in}{}{}{Figure}{\special{language
"Scientific Word";type "GRAPHIC";maintain-aspect-ratio TRUE;display
"USEDEF";valid_file "T";width 2.1607in;height 1.4502in;depth
0.0346in;original-width 7.2218in;original-height 4.8394in;cropleft
"0";croptop "1";cropright "1";cropbottom "0";tempfilename
'PB17ZX00.wmf';tempfile-properties "PR";}}}%
\end{array}%
$

\end{document}